\documentclass[11pt]{article}
\usepackage{fullpage}
\usepackage{verbatim}
\usepackage{amsmath,amsfonts}
\usepackage{amsthm}
\usepackage{url}
\newif\ifabstract
\abstractfalse
\newif\iffull
\ifabstract \fullfalse \else \fulltrue \fi

\usepackage
  [breaklinks,bookmarks,bookmarksnumbered,bookmarksopen,bookmarksopenlevel=2]
  {hyperref}
\hypersetup{pdftitle={De Dictionariis Dynamicis Pauco Spatio Utentibus
  (\emph{lat.} On Dynamic Dictionaries Using Little Space)}}
\hypersetup{pdfauthor={}}

\let\latexcite=\cite
\def\cite{\nolinebreak\latexcite}
\let\latexref=\ref
\def\ref{\nolinebreak\latexref}

{\makeatletter
 \gdef\xxxmark{%
   \expandafter\ifx\csname @mpargs\endcsname\relax 
     \expandafter\ifx\csname @captype\endcsname\relax 
       \marginpar{xxx}
     \else
       xxx 
     \fi
   \else
     xxx 
   \fi}
 \gdef\xxx{\@ifnextchar[\xxx@lab\xxx@nolab}
 \long\gdef\xxx@lab[#1]#2{{\bf [\xxxmark #2 ---{\sc #1}]}}
 \long\gdef\xxx@nolab#1{{\bf [\xxxmark #1]}}
}

\newtheorem{theorem}{Theorem}

\newtheorem{transform}{Transformation}

\newcommand{\eps}{\varepsilon}

\title{\hbox to \hsize{\hss De Dictionariis Dynamicis Pauco Spatio Utentibus\hss}\vskip .6ex
  \large (\emph{lat.} On Dynamic Dictionaries Using Little Space)}

\author{%
        Erik D. Demaine \\
        \small MIT \\
        \small \url{edemaine@mit.edu}
\and
        Friedhelm Meyer auf der Heide \\
        \small U. Paderborn \\
        \small \url{fmadh@upb.de}
\and
        Rasmus Pagh \\
        \small IT U. Copenhagen \\
        \small \url{pagh@itu.dk}
\and
        Mihai P\v{a}tra\c{s}cu \\
        \small MIT \\
        \small \url{mip@mit.edu}
}

\def\and{\end{tabular}\hskip 0.1em plus 0.17fill\begin{tabular}[t]{c}}

\let\epsilon=\varepsilon

\begin{document}

\maketitle

\begin{abstract}
  We develop dynamic dictionaries on the word RAM that use
  asymptotically optimal space, up to constant factors, subject to
  insertions and deletions, and subject to supporting perfect-hashing
  queries and/or membership queries, each operation in constant time
  with high probability.  When supporting only membership queries, we
  attain the optimal space bound of $\Theta(n \lg \frac{u}{n})$ bits,
  where $n$ and $u$ are the sizes of the dictionary and the universe,
  respectively.  Previous dictionaries either did not achieve this
  space bound or had time bounds that were only expected and
  amortized.  When supporting perfect-hashing queries, the optimal
  space bound depends on the range $\{1, 2, \dots, n+t\}$ of hashcodes
  allowed as output.  We prove that the optimal space bound is
  $\Theta(n \lg \lg \frac{u}{n} + n \lg \frac{n}{t+1})$ bits when
  supporting only perfect-hashing queries, and it is $\Theta(n \lg
  \frac{u}{n} + n \lg \frac{n}{t+1})$ bits when also supporting
  membership queries.  All upper bounds are new, as is the $\Omega(n
  \lg \frac{n}{t+1})$ lower bound.
\end{abstract}


\section{Introduction}

The dictionary is one of the most fundamental data-structural problems
in computer science.  In its basic form, a dictionary allows some form
of ``lookup'' on a set $S$ of $n$ objects, and in a dynamic
dictionary, elements can be inserted into and deleted from the
set~$S$.  However, being such a well-studied problem, there are many
variations in the details of what exactly is required of a dictionary,
particularly the lookup operation, and these variations greatly affect
the best possible data structures.  To enable a systematic study, we
introduce a unified view consisting of three possible types of queries
that, in various combinations, capture the most common types of
dictionaries considered in the literature:

\begin{description}
\item[Membership:] Given an element~$x$, is it in the set~$S$?

\item[Retrieval:] Given an element $x$ in the set~$S$, retrieve $r$
  bits of data associated with~$x$.  (The outcome is undefined if $x$
  is not in~$S$.)  The associated data can be set upon insertion or
  with another update operation.  We state constant time bounds for
  these operations, which ignore the $\Theta(r)$ divided by word size
  required to read or write $r$ bits of data.

\item[Perfect hashing:] Given an element $x$ in the set~$S$, return
  the \emph{hashcode} of~$x$.  The data structure assigns to each
  element $x$ in $S$ a unique hashcode in $[n+t]$,%
  \footnote{The notation $[k]$ represents the set $\{0, 1, \dots, k-1\}$.}
  for a specified parameter $t$ (e.g., $t = 0$ or $t = n$).  Hashcodes
  are \emph{stable}: the hashcode of $x$ must remain fixed for the
  duration that $x$ is in~$S$.  (Again the outcome is undefined if $x$
  is not in~$S$.)
\end{description}

Standard hash tables generally support membership and retrieval.  Some
hash tables with open addressing (no chaining) also support perfect
hashing, but the expected running time is superconstant unless $t =
\Omega(n)$.  However, standard hash tables are not particularly space
efficient if $n$ is close to~$u$: they use $O(n)$ words, which is $O(n
\lg u)$ bits for a universe of size~$u$, whereas only $\log_2
\binom{u}{n} = \Theta(n \lg \frac{u}{n})$ bits (assuming $n \leq u/2$)
are required to represent the set~$S$.%
\footnote{Throughout this paper, $\lg x$ denotes $\log_2 (2+x)$,
  which is positive for all $x \geq 0$.}

Any dictionary supporting membership needs at least $\log_2
\binom{u}{n}$ bits of space.  But while such dictionaries are
versatile, they are large, and membership is not always required.  For
example, Chazelle et al.~\cite{chazelle04bloom} explore the idea of a
static dictionary supporting only retrieval, with several applications
related to Bloom filters.  For other data-structural problems, such as
range reporting in one dimension
\cite{mortensen05dyn1d,alstrup01range}, the only known way to get
optimal space bounds is to use a dictionary that supports retrieval
but not membership.  The retrieval operation requires storing the
$r$-bit data associated with each element, for a total of at least $r
n$ bits.  If $r$ is asymptotically less than $\lg \frac{u}{n}$, then
we would like to avoid actually representing the set~$S$.  However, as
we shall see, we still need more than $r n$ bits even in a
retrieval-only dictionary.

Perfect hashing is stronger than retrieval, up to constant factors in
space, because we can simply store an array mapping hashcodes to the
$r$-bit data for each element.  Therefore we focus on developing
dictionaries supporting perfect hashing, and obtain retrieval for
free.  Conversely, lower bounds on retrieval apply to perfect hashing
as well.  Because hashcodes are stable, this approach has the
additional property that the associated data never moves, which can be
useful, e.g.~when the data is large or is stored on disk.

Despite substantial work on dictionaries and perfect hashing (see
Section~\ref{sec:previous}), no dynamic dictionary data structure
supporting any of the three types of queries simultaneously achieves
(1)~constant time bounds with high probability and
(2)~\emph{compactness} in the sense that the space is within a
constant factor of optimal.

\subsection{Our results}

We characterize the optimal space bound, up to constant factors, for a
dynamic dictionary supporting any subset of the three operations,
designing data structures to achieve these bounds and in some cases
improving the lower bound.  To set our results in context, we first
state the two known lower bounds on the space required by a dictionary
data structure.  First, as mentioned above, any dictionary supporting
membership (even static) requires $\Omega(n \lg \frac{u}{n})$ bits of
space, assuming that $n \leq u/2$.  Second, any dictionary supporting
retrieval must satisfy the following recent and strictly weaker lower
bound:

\begin{theorem} {\rm \latexcite{mortensen05dyn1d}} \label{thm:retrieval-lb}
  Any dynamic dictionary supporting retrieval (and therefore any
  dynamic dictionary supporting perfect hashing) requires $\Omega(n
  \lg \lg \frac{u}{n})$ bits of space in expectation, even when the
  associated data is just $r=1$ bit.
\end{theorem}

Surprisingly, for dynamic dictionaries supporting perfect hashing,
this lower bound is neither tight nor subsumed by a stronger lower
bound.  In Section \ref{sec:range-lb}, we prove our main lower-bound
result, which complements Theorem \ref{thm:retrieval-lb} depending on
the value of~$t$:

\begin{theorem} \label{thm:range-lb}
  Any dynamic dictionary supporting perfect hashing with hashcodes in
  $[n+t]$ must use $\Omega(n \lg \frac{n}{t+1})$ bits of space in
  expectation, regardless of the query and update times, assuming that
  $u \geq n + (1+\epsilon) t$ for some constant $\epsilon > 0$.
\end{theorem}

Our main upper-bound result is a dynamic dictionary supporting perfect
hashing that matches the sum of the two lower bounds given by
Theorems~\ref{thm:retrieval-lb} and~\ref{thm:range-lb}.  Specifically,
Section~\ref{sec:ph} proves the following theorem:

\begin{theorem}  \label{thm:ph}
  There is a dynamic dictionary that supports updates and perfect
  hashing with hashcodes in $[n+t]$ (and therefore also retrieval
  queries) in constant time per operation, using $O(n \lg \lg
  \frac{u}{n} + n \lg \frac{n}{t+1})$ bits of space.  The query and
  space complexities are worst-case, while updates are handled in
  constant time with high probability.
\end{theorem}

To establish this upper bound, we find it necessary to also obtain
optimal results for dynamic dictionaries supporting both membership
and perfect hashing.  In Section~\ref{sec:ph+memb}, we find that the
best possible space bound is a sum of two lower bounds in this case as
well:

\begin{theorem} \label{thm:ph+memb}
  There is a dynamic dictionary that supports updates, membership
  queries, and perfect hashing with hashcodes in $[n+t]$ (and
  therefore also retrieval queries) in constant time per operation,
  using $O(n \lg \frac{u}{n} + n \lg\frac{n}{t+1})$ bits of space.
  The query and space complexities are worst-case, while updates are
  handled in constant time with high probability.
\end{theorem}

In the interest of Theorems~\ref{thm:ph} and~\ref{thm:ph+memb}, we
develop a family of \emph{quotient hash functions}. These hash
functions are permutations of the universe; they and their inverses
are computable in constant time given a small-space representation;
and they have natural distributional properties when mapping elements
into buckets. (In contrast, we do not know any hash functions with
these properties and, say, 4-wise independence.)  These hash functions
may be of independent interest.

Table~\ref{summary} summarizes our completed understanding of the
optimal space bounds for dynamic dictionaries supporting updates and
any combination of the three types of queries in constant time with
high probability.  All upper bounds are new, as are the lower bounds
for perfect hashing with or without membership.

\begin{table}
  \centering
  \small
  \tabcolsep=0.15em
  \def\dash{---}
  \ifabstract
    \tabcolsep=0.05em
    \def\quad{\hspace{0.5em}}
    \def\dash{}
  \fi
  \begin{tabular}{|ccccc@{\quad}c@{\quad}l@{\quad}l|}
    \hline
    \multicolumn{5}{|c}{\textbf{Dictionary queries supported}\quad\quad} &  & \multicolumn{1}{c}{\textbf{Optimal space}\quad\quad} & \multicolumn{1}{c|}{\textbf{Reference}}
    \\ \hline
               & & retrieval & &                 & \dash & $\Theta(n \lg \lg \frac{u}{n} + nr)$ & $O$ \iffull in \fi \S\ref{sec:ph}; $\Omega$ \iffull in \fi \cite{mortensen05dyn1d}
    \\
               & & retrieval &+& perfect hashing & \dash & $\Theta(n \lg \lg \frac{u}{n} + n \lg \frac{n}{t+1} + nr)$ & $O$ \iffull in \fi \S\ref{sec:ph}; $\Omega$ \iffull in \fi \S\ref{sec:range-lb}\iffull\ and \else, \fi \cite{mortensen05dyn1d}
    \\
    membership & &           & &                 & \dash & $\Theta(n \lg \frac{u}{n})$ & $O$ \iffull in \fi \S\ref{sec:ph+memb}; $\Omega$ \iffull is standard \else std. \fi
    \\
    membership &+& retrieval & &                 & \dash & $\Theta(n \lg \frac{u}{n} + nr)$ & $O$ \iffull in \fi \S\ref{sec:ph+memb}; $\Omega$ \iffull is standard \else std. \fi
    \\
    membership &+& retrieval &+& perfect hashing & \dash & $\Theta(n \lg \frac{u}{n} + n \lg\frac{n}{t+1} + nr)$ & $O$ \iffull in \fi \S\ref{sec:ph+memb}; $\Omega$ \iffull in \fi \S\ref{sec:range-lb}
    \\ \hline
  \end{tabular}
  \smallskip
  \caption{%
    Optimal space bounds for all types of dynamic dictionaries
    supporting operations in constant time with high probability.
    The upper bounds supporting retrieval without perfect hashing
    can be obtained by substituting $t=n$.
    The $\Theta(n \lg \frac{u}{n})$ bounds assume $n \leq u/2$;
    more precisely, they are $\Theta(\log_2 \binom{u}{n})$.}
  \ifabstract \vspace*{-6ex} \fi
  \label{summary}
\end{table}

\subsection{Previous work}  \label{sec:previous}

There is a huge literature on various types of dictionaries, and we do
not try to discuss it exhaustively. A milestone in the history of
constant-time dictionaries is the realization that the space and query
bounds can be made worst case (construction and updates are still
randomized). This was achieved in the static case by Fredman,
Koml\'{o}s, and Szemer\'{e}di \cite{fredman84dict} with a dictionary
that uses $O(n\lg u)$ bits.  Starting with this work, research on the
dictionary problem evolved in two orthogonal directions: creating
dynamic dictionaries with good update bounds, and reducing the space.

In the dynamic case, the theoretical ideal is to make updates run in
constant time per operation with high probability. After some work,
this was finally achieved by the high-performance dictionaries of
Dietzfelbinger and Meyer auf der Heide \cite{dietzfel90highperf}.
However, this desiderate is usually considered difficult to achieve,
and most dictionary variants that have been developed since then fall
short of it, by having amortized and/or expected time bounds (not with
high probability).

As far as space is concerned, the goal was to get closer to the
information theoretic lower bound of $\log_2 \binom{u}{n}$ bits for
membership.  Brodnik and Munro \cite{brodnik99member} were the first
to solve static membership using $O(n \lg \frac{u}{n})$ bits, which
they later improved to $(1 + o(1)) \log_2 \binom{u}{n}$. Pagh
\cite{pagh01redundancy} solves the static dictionary problem with
space $\log_2 \binom{u}{n}$ plus the best lower-order term known to
date.  For the dynamic problem, the best known result is by Raman and
Rao \cite{raman03succinct}, achieving space $(1+o(1)) \log_2
\binom{u}{n}$.  Unfortunately, in this structure, updates take
constant time amortized and in expectation (not with high
probability).  These shortcomings seem inherent to their technique.

Thus, none of the previous results simultaneously achieve good space
and update bounds, a gap filled by our work.  Another shortcoming of
the previous results lies in the understanding of dynamic dictionaries
supporting perfect hashing.  The dynamic perfect hashing data
structure of Dietzfelbinger et al.~\cite{dietzfel94dph} supports
membership and a weaker form of perfect hashing in which hashcodes are
not stable, though only an amortized constant number of hashcodes
change per update.  This structure achieves a suboptimal space bound
of $O(n \lg u)$ and updates take constant time amortized and in
expectation. No other dictionaries can answer perfect hashing queries
except by associating an explicit hashcode with each element, which
requires $\Theta(n\lg n)$ additional bits.  Our result for membership
and perfect hashing is the first achieving $O(n\lg \frac{u}{n})$
space, even for weak update bounds.  A more fundamental problem is
that all dynamic data structures supporting perfect hashing use
$\Omega(n \lg \frac{u}{n})$ space, even when we do not desire
membership queries so the information theoretic lower bound does not
apply.

Perfect hashing in the static case has been studied intensely, and
with good success. There, it is possible to achieve good bounds with
$t=0$, and this has been the focus of attention. When membership is
required, a data structure using space $(1 + o(1)) \lg \binom{u}{n}$
was finally developed by \cite{raman02perfhash}.  Without membership,
the best known lower bound is $n \log_2 e + \lg \lg u + O(\lg n)$ bits
\cite{fredman84separate}, while the best known data structure uses $n
\log_2 e + \lg \lg u + O(n \frac{(\lg \lg n)^2}{\lg n} + \lg \lg \lg
u)$ bits \cite{hagerup01perfhash}.  Our lower bound depending on $t$
shows that in the dynamic case, even $t = O(n^{1-\eps})$ requires
$\Omega(n\lg n)$ space, making the problem uninteresting.  Thus, we
identify an interesting hysteresis phenomenon, where the dynamic
nature of the problem forces the data structure to remember more
information and use more space.

Retrieval without membership was introduced as ``Bloomier filters'' by
Chazelle et al.~\cite{chazelle04bloom}.  The terminology is by analogy
with the Bloom filter, a static structure supporting approximate
membership (a query we do not consider in this paper) in $\Theta(n \lg
\frac{1}{\epsilon})$ space.  Bloomier filters are static dictionaries
supporting retrieval using $O(n r + \lg \lg u)$ bits of space.  For
dynamic retrieval of $r=1$ bit without membership, Chazelle et
al.~\cite{chazelle04bloom} show that $\Omega(n \lg \lg u)$ bits of
space can be necessary in the case $n^{3+\epsilon} \leq u \leq
2^{n^{O(1)}}$.  Their bound is improved in \cite{mortensen05dyn1d},
giving Theorem~\ref{thm:retrieval-lb}.  On the upper-bound side, the
only previous result is that of \cite{mortensen05dyn1d}: dynamic
perfect hashing for $t = \Theta(n/\lg u)$ using space $O(n \lg\lg u)$.
Our result improves $\lg\lg u$ to $\lg\lg \frac{u}{n}$, and offers the
full tradeoff depending on~$t$.

\subsection{Details of the model}

A few details of the model are implicit throughout this paper.
The model of computation is the Random Access Machine with cells of
$\lg u$ bits (the word RAM).  Because we ignore constant factors,
we assume without loss of generality that $u$, $t$, and $b$
are all exact powers of~$2$.

In dynamic dictionaries supporting perfect hashing, $n$ is not the
current size of the set~$S$, but rather $n$ is a fixed upper bound on
the size of~$S$.  Similarly, $t$ is a fixed parameter.  This
assumption is necessary because of the problem statement: hashcodes
must be stable and the hashcode space is defined in terms of $n$ and~$t$.
This assumption is not necessary for retrieval queries, although we
effectively assume it through our reduction to perfect hashing.
Our results leave open whether a dynamic dictionary supporting
only retrieval can achieve space bounds depending on the current size of
the set $S$ instead of an upper bound~$n$; such a result would in some sense
improve the first row of Table~\ref{summary}.

On the other hand, if we want a dynamic dictionary supporting
membership but not perfect hashing (but still supporting retrieval),
then we can rebuild the data structure whenever $|S|$ changes by a
constant factor, and change the upper bounds $n$ and $t$ then. This
global rebuilding can be deamortized at the cost of increasing space
by a constant factor, using the standard tricks involving two copies
of the data structure with different values of $n$ and~$t$.

Another issue of the model of memory allocation.  We assume that the
dynamic data structure lives in an infinite array of word-length cells.
At any time, the space usage of the data structure is the length
of the shortest prefix of the array containing all nonblank cells.
This model charges appropriately for issues such as external fragmentation
(unlike, say, assuming that the system provides memory-block allocation)
and is easy to implement in practical systems.
See \cite{raman03succinct} for a discussion of this issue.

Finally, we prove that our insertions work in constant time with high
probability, that is, with probability $1-1/n^c$ for any desired
constant~$c > 0$.  Thus, with polynomially small probability, the
bounds might be violated.  For a with-high-probability bound, the data
structure could fail in this low-probability event.  To obtain the
bounds also in expectation and with zero error, we can freeze the
high-performance data structure in this event and fall back to a
simple data structure, e.g., a linked list of any further inserted
elements.  Any operations (queries or deletions) on the old elements
are performed on the high-performance data structure, while any
operations on new elements (e.g., insertions) are performed on the
simple data structure.  The bounds hold in expectation provided that
the data structure is used for only a polynomial amount of time.

\section{Quotient Hash Functions} \label{sec:quothash}

We define a quotient hash function in terms of three parameters: the
universe size $u$, the number of buckets $b$, and an upper bound $n$
on the size of the sets of interest. A quotient hash function is
simply a bijective function $h : [u] \to [b] \times [\frac{u}{b}]$.
We interpret the first output as a bucket, and the second output as a
``quotient'' which, together with the bucket, uniquely identifies the
element. We write $h(x)_1$ and $h(x)_2$ when we want to refer to
individual outputs of $h$.

We are interested in sets of elements $S \subset [u]$ with $|S| \le
n$. For such a set $S$ and an element $x$, define $\mathcal{B}_h(S, x)
= \{ y \in S \mid h(y)_1 = h(x)_1 \}$, i.e.~the set of elements mapped
to the same bucket as $x$. For a threshold $t$, define
$\mathcal{C}_h(S, t) = \{ x \in S \mid \#\mathcal{B}_h(S,x) \ge t \}$,
i.e.~the set of elements which map to buckets containing at least $t$
elements. These are elements that ``collide'' beyond the allowable
threshold.

\begin{theorem} \label{thm:quothash}
There is an absolute constant $\alpha < 1$ such that for any $u, n$
and $b$, there exists a family of quotient hash functions $\mathcal{H}
= \{ h : [u] \to [b] \times [\frac{u}{b}] \}$ satisfying:

\begin{itemize}
\item an $h \in \mathcal{H}$ can be represented in $O(n^{\alpha})$
  space and sampled in $O(n^{\alpha})$ time.

\item $h$ and $h^{-1}$ can be evaluated in constant time on a RAM;

\item for any fixed $S \subset [u], |S| \le n$ and any $\delta < 1$,
  the following holds with high probability over the choice of $h$:
\[ 
\left\{ \begin{array}{l@{\qquad}rcl}
  \textrm{if $b \ge n$,} 
  & \#\mathcal{C}_h(S, 2) &\le& 2\frac{n^2}{b} + n^{\alpha} \\
  \textrm{if $b < n$,} 
  & \#\mathcal{C}_h(S, (1 + \delta) \frac{n}{b} + 1) &\le& 
    2n e^{-\delta^2 n / (3b)} + n^{\alpha}
\end{array} \right.
\]
\end{itemize}
\end{theorem}

It is easy to get an intuitive understanding of these bounds. In the
case $b \ge n$, the expected number of collisions generated by
universal hashing (2-independent hashing) would be
$\frac{n^2}{b}$. For $b < n$, we can compare against a highly
independent hash function. Then, the expected number of elements that
land in overflowing buckets is $n e^{-\delta^2 n / (3b)}$, by a simple
Chernoff bound. Our family matches these two bounds, up to a constant
factor and an additive error term of $O(n^{\alpha})$, which are both
negligible for our purposes. The advantage of our hash family is
two-fold. First, it gives quotient hash functions, which is essential
for our data structure. Second, the number of overflowing elements is
guaranteed with high probability, not just in expectation.

\subsection{Permutation hash functions}

It is useful to relate the concept of quotient hash functions to
another concept, namely permutation hash functions. Such functions are
bijections from the universe to itself. We call a family of
permutations \emph{$k$-independent} if, for any input set $S \subset
[u]$ with at most $k$ elements, the output of a randomly drawn hash
function applied to the elements of $S$ is indistinguishable from the
output of a truly random permutation.

It is easy to construct 2-independent permutations. A standard
construction for universal hashing is to map $x \mapsto ax + b$, where
$a$ and $b$ are random, and all elements come from a field
$\mathbb{Z}_u$. Maintaining the same family with the restriction $a
\ne 0$ gives 2-independent permutations. Unfortunately, it is not
known how to construct good $k$-wise independent families, for $k >
3$. In fact, small families are not even known to exist. If one is
satisfied with almost $k$-wise independence, small families can be
constructed (see \cite{kaplan05permhash} and citations therein), but
it is not known how to achieve constant evaluation time.

A permutation hash function can be converted trivially into a quotient
hash function: the bucket is given by some $\lg b$ bits of the output,
and the quotient is the rest. To achieve bounds similar to that of
Theorem \ref{thm:quothash}, all we need is a family of permutations
with sufficiently high independence, but as mentioned already, we do
not know how to construct one. The trick behind Theorem
\ref{thm:quothash} is to recognize that permutation families, though
theoretically clean and interesting, are stronger than what we
need. We can instead get away with slightly weaker concentration
guarantees. One should note that the family of hash functions we
propose does not have any $k$-independence guarantee.

\subsection{Construction of the hash family}

It will be convenient to assume that $u \le n^c$, for some constant
$c$. If this does not hold, we can reduce the universe to $n^c$, for
some big enough $c$, by the following:

\begin{transform} 
Apply a random 2-independent permutation on the original universe.
Keep only the first $c\lg n$ bits of the result, and make the rest
part of the quotient.
\end{transform}

The expected number of collisions generated by keeping only $c\lg n$
bits is bounded by $n^{2-c}$. Thus, by choosing a large enough
constant $c$, we can avoid any collision with the required
high-probability guarantee.

{From} now on, it will be convenient to interpret the universe as a
two-dimensional table, with $n^{3/4}$ columns, and $\frac{u}{n^{3/4}}$
rows. The plan is to use this column structure as a means of
generating independence. Imagine a hash function that generates few
collisions in expectation, but not necessarily with high
probability. However, we can apply a \emph{different} random hash
function inside each column. The expectation is unchanged, but now
Chernoff bounds can be used to show that we are close to the
expectation with high probability, because the behavior of each column
is independent. 

However, to put this plan into action, we need to guarantee that the
elements of $S$ are spread rather uniformly across columns.  We do
this by applying a random circular shift to each row.

\begin{transform}
Consider a highly independent hash function mapping row numbers to
$[n^{3/4}]$. Inside each row, apply a circular shift by the hash
function of that row.
\end{transform}

Note that the number of rows can be pretty large (larger than $n$), so
we cannot afford a truly random shift for each row. However, the
number of rows is polynomial, and we can use Siegel's family of highly
independent hash functions \cite{siegel04hash} to generate highly
independent shifts. These hash functions can be represented in space
$n^{1 - \Omega(1)}$, take constant time to evaluate, and are
$n^{\Omega(1)}$-independent with high probability.

Because shifts are uniform, the expected number of items in each
column is $\frac{|S|}{n^{3/4}} \le n^{1/4}$. We can apply Chernoff
bounds for random variables with limited independence to show that a
column does not have more than $n^{1/4} + n^{1/5}$ elements with very
high probability. Specifically, we apply \cite[Theorem
5.I]{schmidt95chernoff}. Let $X_i$ be the indicator random variable,
specifying whether an element from $S$ is mapped to our column of
interest in row $i$. Then $\mu = E[\sum X_i] = n^{1/4}$. We are
interested in the event $\sum X_i \ge \mu + n^{1/5}$. The theorem
guarantees that, for independence $k = O\big(\frac{(n^{1/5})^2}{\mu}\big) =
O(n^{0.15})$, the probability of this event is $e^{-\Omega(k)}$. Because
Siegel's hash functions give $k = n^{\Omega(1)}$, this event happens
with exponentially small probability.

In the remainder, we consider only $n^{1/4}$ elements of $S$ from each
column. In the worst case, all the excess values from $S$ will be
mapped to buckets that are already full due to these ``normal''
elements. However, the total number of excess elements is at most
$n^{3/4} \cdot n^{1/5} = n^{0.95}$ with high probability, and this is
swallowed by our $n^{\alpha}$ error term.

\paragraph{The case $b \ge n$.}
Remember that, in this case, our goal is to get close to the
collisions generated by a 2-independent permutation, but with high
probability. As explained above, we can achieve this effect through
column independence.

\begin{transform}  \label{transf:3}
Apply a 2-universal permutation inside each column. For each column,
the permutation is chosen independently at random.
\end{transform}

We have only $n^{3/4}$ columns, and a 2-universal permutation takes
$O(\lg n)$ bits to represent, so we can afford to store all these
permutations. To complete the construction, break each column into
$\frac{b}{n^{3/4}}$ equal-sized buckets. The position within a bucket
is thrown away as part of the quotient.

Using a classic Chernoff bound, we show that imposing the bucket
granularity does not generate too many collisions. Let $X_i$ be the
number of elements in column $i$ that are mapped to the same table cell as
some other element. Because we look at only $n^{1/4}$ elements in each
column, $X_i \in [0, n^{1/4}]$. Let $\mu = E[\sum X_i]$; by linearity
of expectation, $\mu \le \frac{n^2}{b}$. By the Chernoff bound,
$\Pr[\sum X_i \ge 2\mu + n^{3/4}] \le \exp(-\Omega(\frac{\mu +
n^{3/4}}{n^{1/4}})) = e^{-\Omega(\sqrt{n})}$. Thus, with exponentially
high probability, the number of collisions is bounded by $\frac{2
n^2}{b} + n^{3/4}$. This completes the analysis.

\paragraph{The case $b < n$.}
Ideally, each bucket should contain $\frac{n}{b}$ elements. We are
interested in buckets of size exceeding $(1 + \delta) \frac{n}{b}$,
and want to bound the number of elements in such buckets close to the
expected number for a highly independent permutation.

As explained already, we do not know any family of highly independent
permutations that can be represented with small space and evaluated
efficiently. Instead, we will revert to the brute-force solution of
representing truly random permutations.  To use this idea and keep the
space small, we need two tricks. The first trick is to generate and
store fewer permutations than columns. It turns out that re-using the
same permutation for multiple columns still gives enough independence.

Note, however, that we cannot even afford to store a random
permutation inside a single column, because columns might have more than
$n$ elements. However, we can reduce columns to size $\sqrt{n}$ as
follows. Use Transformation \ref{transf:3} with $b' = n^{5/4}$. This
puts elements into $n^{5/4}$ \emph{first-order buckets} ($\sqrt{n}$
buckets per column), with a negligible number of collisions. Thus, we
can now work at the granularity of first-order buckets, and ignore the
index within a bucket of an element.

\begin{transform}
Group columns into $n^{1/4}$ equal-sized groups. For each group,
generate a random permutation on $\sqrt{n}$ positions, and apply it to
the first-order buckets inside each column of the group.
\end{transform}

The space required to represent the permutations is $O(n^{3/4} \lg n)$
bits. To complete the construction, we just break columns into
$\frac{b}{n^{3/4}}$ equal-sized buckets. In other words,
$\frac{n^{5/4}}{b}$ first-order buckets are grouped into an output
bucket. If $b \le n^{3/4}$, we are already done, since it suffices to
know elements are well distributed in columns.

We begin by analyzing the probability that a fixed element $x$ ends up
in an overflowing bucket. The events that other elements end up in the
same bucket are negatively correlated (because we have permutations).
Thus, we can upper bound the probability by assuming that the other
$n^{1/4}$ elements can be independently mapped to $x$'s bucket with
probability $\frac{n^{3/4}}{b}$. Then, by the Chernoff bound, the
probability that $(1 + \delta) \frac{n}{b}$ other elements get mapped
to $x$'s bucket is at most $e^{-\delta^2 n / (3b)}$. This is the
probability $x$ that is in an overflowing bucket.

Now let $X_i$ be the number of elements that overflow in column group
$i$. We have $X_i \in [0, n^{3/4}]$ and $E[\sum X_i] = \mu \le n
e^{-\delta^2 n / (3b)}$. Then, by the Chernoff bound, $\Pr[ \sum X_i
\ge 2 \mu + n^{\alpha}] \le \exp(-\Omega( \frac{\mu + n^{\alpha}}
{n^{3/4}} )) = e^{-n^{\Omega(1)}}$ if $\alpha > \frac{3}{4}$. This
completes the analysis.

\subsection{Coping with a dynamic set}

We now discuss an important subtlety in the probabilistic analysis
needed by our data structure. In general, we are dealing with a
dynamic set $S$, and mapping it through a quotient hash function. The
analysis above shows that, for any fixed $S$, the number of overflow
elements is small. However, imagine an element $x$ that becomes an
overflow element at the time of insertion. Then, as some other element $y$
is deleted, $x$ might not be in an overflow condition in the old set, but
our data structure has already handled it as an overflow element. 
Thus, it does not suffice to look at the overflow conditions for a
fixed set $S$. The relevant quantity is the number of elements that
caused an overflow at the time of insertion.

Fortunately, the same bounds from above hold, were $n$ is an upper
bound on the size of $S$. This follows from the same analysis, but
interpreted in a more subtle way. The probability that an insertion
causes an overflow is unchanged. By linearity of expectation (over the
elements currently in the set $S$), the expected number of elements
that cause an overflow when inserted is the same as the expected
bounds from above. But whether an element causes an overflow depends
only on the random coins pertaining to the column of the element.
Thus, by the same independence considerations as above, this
expectation is matched with high probability.

\section{Solution for Membership and Perfect Hashing} \label{sec:ph+memb}

There are two easy cases.
First, if $u = \Omega(n^{1.5})$, then the space bound is $\Theta(n \lg u)$.
In this case, a solution with hashcode range exactly $[n]$ can be obtained by
using a high-performance dictionary \cite{dietzfel90highperf}. We
store an explicit hashcode as the data associated with each value, and
maintain a list of free hashcodes. This takes $O(n\lg n + n \lg u) =
O(n\lg u)$ bits.  Second, if $t = O(n^{\alpha})$, for $\alpha < 1$, then
the space bound is $\Theta(n \lg n)$. Because $u = O(n^{1.5})$, we can use
the same brute-force solution. In the remaining cases, we can assume $t
\le \frac{n^2}{u}$ (we are always free to decrease $t$), so that the
space bound is dominated by $\Theta(n \lg\frac{n}{t})$.

The data structure is composed of three levels. An element is inserted
into the first level that can handle it. The first-level filter
outputs hashcodes in the range $[n + \frac{t}{3}]$, and handles most
elements of $S$: at most $c_1 t$ elements (for a constant $c_1 \le
\frac{1}{3}$ to be determined) are passed on to the second level, with
high probability. The goal of the second-level filter is to handle all
but $O(\frac{n}{\lg n})$ elements with high probability. If $c_1 t \le
\frac{n}{\lg n}$, this filter is not used. Otherwise, we use this
filter, which outputs hashcodes in the range $[\frac{t}{3}]$.
Finally, the third level is just a brute-force solution using a
high-performance dictionary. Because it needs to handle only $\min \{
O(\frac{n}{\lg n}), c_1 t \}$ elements, the output range can be
$[\frac{t}{3}]$ and the space is $O(n)$ bits. This dictionary can
always be made to work with high probability in $n$ (e.g.~by inserting
dummy elements up to $\Omega(\sqrt{n})$ values).

A query tries to locate the element in all three levels. Because all
levels can answer membership queries, we know when we've located an
element, and we can just obtain a hashcode from the appropriate
level. Similarly, deletion just removes the element from the
appropriate level.

\subsection{The first-level filter}

Let $\mu = c_2 (\frac{n}{t})^3$, for a constant $c_2$ to be
determined. We use a quotient hash function mapping the universe into
$b = \frac{n}{\mu}$ buckets. Then, we expect $\mu$ elements per
bucket, but we will allow for an additional $\mu^{2/3}$ elements. By
Theorem \ref{thm:quothash}, the number of elements that overflow is
with high probability at most $n e^{-\Omega(\sqrt[3]{\mu})} +
n^{\alpha}$. For big enough $c_2$, this is at most $\frac{c_1}{2} t$
(remember that we are in the case when $n^{\alpha}$ is negligible).

Now we describe how to handle the elements inside each bucket. For
each bucket, we have a hashcode space of $[\mu + \mu^{2/3}]$.
Then, the code space used by the first-level filter is $n +
\frac{n}{\sqrt[3] \mu} \le n + \frac{t}{3}$ for big enough $c_2$.  We
use a high-performance dictionary inside each bucket, which stores
hashcodes as associated data. We also store a list of free hashcodes
to facilitate insertions. To analyze the space, observe that a hashcode
takes only $O(\lg \frac{n}{t})$ bits to represent. In addition,
the high-performance dictionary need only store the \emph{quotient} of
an element. Indeed, the element is uniquely identified by the
quotient and the bucket, so to distinguish between the elements in a
bucket we only need a dictionary on the quotients. Thus, we need
$O(\lg \frac{u\mu}{n}) = O(\lg \frac{u}{n} + \lg \frac{n}{t})$ bits
per element.

The last detail we need to handle is what happens when an insertion in
the bucket's dictionary fails. This happens with probability
$\mu^{-c_3}$ for each insertion, where $c_3$ is any desired
constant. We can handle a failed insertion by simply passing the
element to the second level. The expected number of elements whose
insertion at the first level failed is $n \mu^{-c_3} \le \frac{c_1}{4}
t$ for big enough $c_3$. Since we can assume $t = \Omega(n^{5/6})$, we
have $\mu = O(\sqrt{n})$ and $b = \Omega(\sqrt{n})$. This means we
have $\Omega(\sqrt{n})$ dictionaries, which use independent random
coins. Thus, a Chernoff bound guarantees that we are not within twice
this expectation with probability at most $e^{-\Omega(t / \sqrt{n})} =
e^{-n^{\Omega(1)}}$ because $t = \Omega(n^{\alpha})$.  Thus, at most
$c_1 t$ elements in total are passed to the second level with high
probability.

\subsection{The second-level filter}

We first observe that this filter is used only when $\lg \frac{u}{n} =
O(\sqrt[4]{\lg n})$. Indeed, $t \le \frac{n^2}{u}$, so when $\lg
\frac{u}{n} = \Omega(\sqrt[4]{\lg n})$, we have $t = o(\frac{n}{\lg
n})$, and we can skip directly to the third level.

We use a quotient hash function mapping the universe to $b =
\frac{c_1 t}{\sqrt[4]{\lg n}}$ buckets. We allow each bucket to
contain up to $2\sqrt[4]{\lg n}$ elements; overflow elements are
passed to the third level. By Theorem \ref{thm:quothash}, at most $n /
2^{\Omega(\sqrt[4]{\lg n})} = o(n / \lg n)$ elements are passed to the
third level, with high probability.

Because buckets contain $O(\sqrt[4]{\lg n})$ elements of
$O(\sqrt[4]{\lg n})$ bits each, we can use word-packing tricks to
handle buckets in constant time. However, the main challenge is space,
not time. Observe that we can afford only $O(\lg \frac{u}{n})$ bits
per element, which can be much smaller than $O(\sqrt[4]{\lg n})$. This
means that we cannot even store a permutation of the elements inside a
bucket. In particular, it is information-theoretically impossible
even to store the elements of a bucket in an arbitrary order!

Coping with this challenge requires a rather complex solution: we
employ $O(\lg\lg n)$ levels of filters and permutation hashing inside
each bucket. Let us describe the level-$i$ filter inside a
bucket. First, we apply a random permutation to the bucket universe
(the quotient of the elements inside the bucket). Then, the filter
breaks the universe into $\frac{c_4 \sqrt[4]{\lg n}}{2^i}$ equal-sized
\emph{tiles}. The filter consists of an array with one position per
tile. Such a position could either be empty, or it stores the index
within the tile of an element mapped to that tile (which is a quotient
induced by the permutation at this level). Observe that the size of
the tiles doubles for each new level, so the number of entries in the
filter array halves. In total, we use $h = \frac{1}{8} \lg\lg n$
filters, so that the number of tiles in any filter is
$\Omega(\sqrt[8]{\lg n})$.

Conceptually, an insertion traverses the filters sequentially starting
with $i = 0$. It applies permutation $i$ to the element, and checks
whether the resulting tile is empty. If so, it stores the element in
that tile; otherwise, it continues to the next level. Elements that
cannot be mapped in any of the $h$ levels are passed on to the third
level of our big data structure. A deletion simply removes the element
from the level where it is stored. A perfect-hash query returns the
identifier of the tile where the element is stored. Because the number
of tiles decreases geometrically, we use less than $2 c_4 \sqrt[4]{\lg
n}$ hashcodes per bucket. We have $\frac{c_1 t}{\sqrt[4]{\lg n}}$
buckets in total and we can make $c_1$ as small as we want, so the
total number of hashcodes can be made at most $\frac{t}{3}$.

We now analyze the space needed by this construction. Observe that the
size of the bucket universe is $v = u \cdot \frac{\sqrt[4]{\lg n}}{c_1
t}$. Thus, at the first level, the filter requires $\lg \frac{v}{c_4
\sqrt[4]{\lg n}}$ bits to store an index within each tile. At each
consecutive level, the number of bits per tile increases by one
(because tiles double in size), but the number of tiles halves. Thus,
the total space is dominated by the first level, and it is $O(\lg
\frac{u}{t}) = O(\lg \frac{u}{n} + \lg \frac{n}{t})$ bits per element.

\paragraph{Bounding the unfiltered elements.}
We now want to bound the expected number of elements in a bucket that
are not handled by any of these filters. In doing so, we assume that the
permutation hash functions at each level are truly random; this will
be justified later. We assume that, at level~$i$, there are at most
$\frac{2\sqrt[4]{\lg n}}{2^{i c_5}}$ elements that haven't been
handled by previous levels (for $c_5$ to be determined). When
analyzing filter $i$, there is a probability that this condition will
not be met at level $i+1$. If this happens, we just assume
pessimistically that all remaining elements are rejected by all
filters. However, the probability of this event will be small enough,
so the total expectation of the number of rejected elements will be
small.

At level $i$ there are $\frac{c_4 \sqrt[4]{\lg n}}{2^i}$ tiles, so an
element has probability at most $\frac{2}{c_4} 2^{-(c_5 - 1) i}$ of
conflicting with another element. Then the expected number of
conflicting elements is $\mu = \frac{4\sqrt[4]{\lg n}}{c_4} 2^{-(2c_5
- 1) i}$. Conflicts are negatively correlated, so we can apply a
Chernoff bound on the number of elements not handled at this level. We
want to analyze the case when there are more than $(1+\delta)\mu =
\frac{2\sqrt[4]{\lg n}}{2^{c_5(i+1)}}$ unhandled elements (which
invalidates our assumption at level $i+1$). We have $1+\delta =
\frac{c_4}{2^{c_5 + 1}} 2^{(c_5 - 1) i}$. By Chernoff bounds, the
probability of this event is at most $\left( e^{\delta}
(1+\delta)^{-(1+\delta)} \right)^{\mu} \le \left( \frac{e}{1+\delta}
\right)^{(1+\delta) \mu}$.  We now distinguish three ranges for $i$:

\begin{itemize}
\item As long as $(1+\delta) \mu \ge \sqrt[8]{\lg n}$, observe that,
  for any~$c_5$, $c_4$ can be chosen large enough so that $1 + \delta
  > 2e$, so the probability of the bad event is
  $2^{-\Omega(\sqrt[8]{\lg n})}$. Thus, the contribution of this bad
  event to the total expectation is $o(\frac{1}{\lg n})$, even summed
  over all levels.

\item As long as $(1+\delta) \mu \ge 16$, the probability is bounded by
  $(\frac{e}{1+\delta})^{16}$. Note that since $(1+\delta) \mu \le
  \sqrt[8]{\lg n}$, we have $1+\delta = \Omega(\sqrt[8]{\lg
  n})$. Then, the failure probability is $O(\frac{1}{\lg^2 n})$, so
  the contribution to the total expectation is $o(\frac{1}{\lg n})$,
  even summing over all levels.

\item If $(1+\delta) \mu \le 16$, we only have to handle $O(1)$
  elements. The probability of any conflict is inversely proportional
  to the number of tiles, which is $\Omega(\sqrt[8]{\lg n})$. The
  probability of a persistent conflict decreases exponentially with
  the number of levels, so after $O(1)$ levels, we expect
  $o(\frac{1}{\lg n})$ conflicts. For large enough $c_5$, $(1+\delta)
  \mu$ decreases sufficiently rapidly that we will actually have the
  required $O(1)$ levels in this range.
\end{itemize}

\paragraph{Implementation details.}
We now discuss how to implement the conceptual ideas from above, and
make operations take constant time. Note that $\lg v = O(\sqrt[4]{\lg
n})$. Thus, the output of $O(\lg\lg n)$ permutation functions takes
$O(\sqrt[4]{\lg n} \cdot \lg\lg n)$ bits, and easily fits in a
word. For now, assume that we can evaluate the output of all these
functions in constant time. The description of an entire bucket takes
$O(\sqrt[4]{\lg n} \cdot \lg \frac{u}{n}) = O(\sqrt{\lg n})$ bits, so
a bucket can also be manipulated as a word. Then, updates and queries
of a bucket are functions on a range of $O(\sqrt[4]{\lg n} \cdot
\lg\lg n)$ bits for the argument, and $O(\sqrt{\lg n})$ bits for the
bucket representation. So they can be implemented in constant time
using a lookup table of size $\exp(O(\lg^{3/4} n \cdot \lg\lg n))
\cdot O(\lg n) = n^{o(1)}$ bits.

It remains to implement the random permutations. We can simply
generate $O(\lg\lg n)$ random permutations on the universe $[v]$, and
create a lookup table with the packed output of all permutations,
applied to each possible input. This takes space $\exp(O(\sqrt[4]{\lg
n})) \cdot O(\lg n)$ bits. We cannot afford such a collection of
random permutations for each bucket, but we can re-use just a few random
permutations. In particular, we can generate $n^{2/3}$ independent
lookup tables, and use each one for an equal share of the buckets.
Then, the number of elements that get passed to the third level
because they are not handled by any filter inside their bucket is the
sum of $n^{2/3}$ independent components. By the Chernoff bound, the
expectation (analyzed above) is exceeded by a constant factor
only with exponentially small probability.

\section{Solution for Perfect Hashing}  \label{sec:ph}

The data structure supporting perfect hashing but not membership
consists of one quotient hash function, selected from the family of
Theorem~\ref{thm:quothash}, and two instances of the data structure of
Theorem~\ref{thm:ph+memb} supporting perfect hashing and
membership. The quotient hash function divides the universe into $b$
buckets, and we set $b = \frac{c \, n^2}{t+1}$ for a constant $c \geq
1$ to be determined.

The first data structure supporting perfect hashing and membership
stores the set $B$ of buckets occupied by at least one element
of~$S$. An entry in $B$ effectively represents an element of $S$ that
is mapped to that bucket. However, we have no way of knowing the exact
element. The second data structure supporting perfect hashing and
membership stores the additional elements of $S$, which at the time of
insertion were mapped to a bucket already in $B$.

Insertions check whether the bucket containing the element is in
$B$. If not, we insert it.  Otherwise, we insert the element into the
second data structure.  Deletions proceed in the reverse order.
First, we check whether the element is listed in the second data
structure, in which case we delete it from there.  Otherwise, we
delete the bucket containing the element from the first data
structure.

The range of the first perfect hash function should be $[n +
\frac{t}{2}]$. For the second one, it should be $[\frac{t}{2}]$; we
show below that this is sufficient with high probability. Thus, we use
$[n+t]$ distinct hashcodes in total.  To perform a query, we first
check whether the element is listed in the second data structure.  If
it is, we return the label reported by that data structure (offset by
$n + \frac{t}{2}$ to avoid the hashcodes from the first data
structure).  Otherwise, because we assume that the element is in~$S$,
it must be represented by the first data structure.  Thus, we compute
the bucket assigned to the element by the quotient hash function, look
up that bucket in the first data structure, and return its label.

It remains to analyze the space requirement. We are always free to
reduce $t$, so we can assume $t = O(n / \lg \frac{u}{n})$, simplifying
our space bound to $O(n\lg \frac{n}{t+1})$. Because $|B| \le n$, the
first data structure needs space $O(\lg \binom{b}{n} + n \lg
\frac{n}{t/2+1}) = O(n \lg \frac{b}{n} + n\lg \frac{n}{t+1}) = O(n \lg
\frac{n}{t+1})$, which is within the desired bound.

Because $b \ge n$, our family of hash functions guarantees that, with
high probability, the number of elements of $S$ that were mapped to a
nonempty bucket at the time of their insertion is at most $\frac{2
n^2}{b} + n^\alpha = \frac{2 (t+1)}{c} + n^\alpha$. If $n^{\alpha} <
\frac{t}{8}$, this is at most $\frac{t}{4}$ for sufficiently large
$c$. If $t = O(n^{\alpha})$, we can use a brute-force solution: first,
construct a perfect hashing structure with $t = n$ (this is possible
through the previous case); then, relabel the used positions in the
$[2n]$ range to a minimal range of $[n]$, using $O(n\lg n)$ memory
bits.

Given this bound on the number of elements in the second structure,
note that the number of hashcodes allowed ($t/2$) is double the
number of elements. Thus the space required by the second structure is
$O(\lg \binom{u}{t/4} + t) = O(t \lg \frac{u}{t}) =
O(\frac{n}{\lg(u/n)} \lg \frac{u}{n / \lg(u/n)}) = O(n)$.

\section{Lower Bound for Perfect Hashing} \label{sec:range-lb}

This section proves Theorem \ref{thm:range-lb} assuming $u \ge 2n$. We
defer case of smaller $u$ to the full version. Our lower bound
considers the dynamic set $S$ which is initially $\{ n+1, \dots, 2n
\}$ and is transformed through insertions and deletions into $\{ 1,
\dots, n \}$. More precisely, we consider $\frac{n}{2t}$ stages. In
stage $i$, we pick a random subset $D_i \subseteq S \cap \{ n+1,
\dots, 2n \}$, of cardinality~$2t$. Then, we delete the elements in
$D_i$, and we insert elements $I_i = \{ (i-1)2t + 1, \dots, i\cdot 2t
\}$. Note than, in the end, the set is $\{1, \dots, n\}$. By the easy
direction of Yao's minimax principle, we can fix the random bits of
the data structure, such that it uses the same expected space over the
input distribution.

Our strategy is to argue that the data structure needs to remember a
lot of information about the history, i.e.~there is large hysteresis
in the output of the perfect hash function. Intuitively, the $2t$
elements inserted in each stage need to be mapped to only $3t$
positions in the range: the $t$ positions free at the beginning of the
stage, and the $2t$ positions freed by the recent deletes. These free
positions are quite random, because we deleted random elements. Thus,
this choice is very constrained, and the data structure needs to
remember the constraints.

Let $h$ be a function mapping each element in $[2n]$ to the hashcode
it was assigned; this is well defined, because each element is assigned
a hashcode exactly once (though for different intervals of time). We
argue that the vector of sets $(h(I_1), \dots, h(I_{n/2t}))$ has
entropy $\Omega(n\lg\frac{n}{t})$. One can recover this vector by
querying the final state of the data structure, so the space lower
bound follows.

We first break up the entropy of the vector by: $H(h(I_1), \dots,
h(I_{n/2t})) = \sum_j H(h(I_j) \mid h(I_1), \dots, h(I_{j-1}))$.  Now
observe that the only randomness up to stage $j$ is in the choices of
$D_1, \dots, D_{j-1}$.  In other words, $D_1, \dots, D_{j-1}$
determine $h(I_1), \dots, h(I_{j-1})$. Then, $H(h(I_1), \dots,
h(I_{n/2t})) \ge \sum_j H(h(I_j) \mid D_1, \dots, D_{j-1})$. To
alleviate notation, let $D_{<j}$ be the vector $(D_1, \dots,
D_{j-1})$.

Now we lower bound each term of the sum. Let $F_j$ be the set of free
positions in the range at the beginning of stage $j$. Because we made
the data structure deterministic, $F_j$ is fixed by conditioning on
$D_{<j}$. Because $I_j$ can be mapped to free positions only after
$D_j$ is deleted, we find that $h(I_j) \subset F_j \cup h(D_j)$. Note
that $|h(I_j)| = 2t$, but $|F_j| = t$. Thus, $|h(D_j) \setminus
h(I_j)| \le t$.

Now we argue that the entropy of $h(D_j)$ is large. Indeed, $D_j$ is
chosen randomly from $S \cap \{n+1, \dots, 2n\}$, a set of cardinality
$n - 2t(j-1)$. Conditioned on $D_{<j}$, the set $S \cap \{n+1, \dots,
2n\}$ is fixed, so its image through $h$ is fixed. Then, choosing
$D_j$ randomly is equivalent to choosing $h(D_j)$ randomly from a
fixed set of cardinality $n - 2t(j-1)$. So $H(h(D_j) \mid D_{<j}) =
\lg \binom{n-2t(j-1)}{2t}$. Now consider $h(D_j) \setminus
h(I_j)$. This is a set of cardinality at most $t$ from the same set of
$n-2t(j-1)$ positions. Thus, $H(h(D_j) \setminus h(I_j) \mid D_{<j})
\le \lg \binom{n-2t(j-1)}{t} + t$.

Using $H(a,b) \le H(a) + H(b)$, we have
  \vspace{-.1cm}
\[ H(h(D_j) \mid D_{<j}) \le H(h(D_j) \cap h(I_j) \mid D_{<j}) +
  H(h(D_j) \setminus h(I_j) \mid D_{<j}). \]

Of course, $H(h(I_j) \mid D_{<j}) \ge H(h(I_j) \cap h(D_j) \mid
D_{<j})$. This implies
  \vspace{-.1cm}
\begin{eqnarray*}
H(h(I_j) \mid D_{<j})
  &\ge& H(h(D_j) \mid D_{<j}) - H(h(D_j) \setminus h(I_j) \mid D_{<j}) \\
  &\ge& \lg \binom{n-2t(j-1)}{2t} - \lg \binom{n-2t(j-1)}{t} - t.
\end{eqnarray*}

Using $\binom{a}{b} / \binom{a}{c} = \binom{a-c}{b}$, we have
$H(h(I_j) \mid D_{<j}) \ge \lg \binom{n-t(2j-1)}{2t} - t$. For $j \le
\frac{n}{3t}$, we have $H(h(I_j) \mid D_{<j}) = \Omega(t \lg
\frac{n}{t})$. We finally obtain $H(h(I_1), \dots, h(I_j)) = \Omega(n
\lg \frac{n}{t})$, concluding the proof.

\small

\vspace{-.2cm}
\paragraph{Acknowledgement.}
Part of this work was done at the Oberwolfach Meeting on
Complexity Theory, July 2005.

\vspace{-.4cm}


\bibliographystyle{plain}
\bibliography{../general}

\begin{thebibliography}{10}

\bibitem{alstrup01range}
Stephen Alstrup, Gerth Brodal, and Theis Rauhe.
\newblock Optimal static range reporting in one dimension.
\newblock In {\em Proc. 33rd ACM Symposium on Theory of Computing (STOC)},
  pages 476--482, 2001.

\bibitem{brodnik99member}
Andrej Brodnik and J.~Ian Munro.
\newblock Membership in constant time and almost-minimum space.
\newblock {\em SIAM Journal on Computing}, 28(5):1627--1640, 1999.
\newblock See also ESA'94.

\bibitem{chazelle04bloom}
Bernard Chazelle, Joe Kilian, Ronitt Rubinfeld, and Ayellet Tal.
\newblock The {B}loomier filter: an efficient data structure for static support
  lookup tables.
\newblock In {\em Proc. 15th ACM/SIAM Symposium on Discrete Algorithms (SODA)},
  pages 30--39, 2004.

\bibitem{dietzfel90highperf}
Martin Dietzfelbinger and Friedhelm~Meyer auf~der Heide.
\newblock A new universal class of hash functions and dynamic hashing in real
  time.
\newblock In {\em Proc 17th International Colloquium on Automata, Languages and
  Programming (ICALP)}, pages 6--19, 1990.

\bibitem{dietzfel94dph}
Martin Dietzfelbinger, Anna Karlin, Kurt Mehlhorn, Friedhelm Meyer Auf~Der
  Heide, Hans Rohnert, and Robert~E. Tarjan.
\newblock Dynamic perfect hashing: Upper and lower bounds.
\newblock {\em SIAM Journal on Computing}, 23(4):738--761, 1994.
\newblock See also FOCS'88.

\bibitem{fredman84separate}
Michael~L. Fredman and J\'{a}nos Koml\'{o}s.
\newblock On the size of separating systems and families of perfect hash
  functions.
\newblock {\em SIAM Journal on Algebraic and Discrete Methods}, 5(1):61--68,
  1984.

\bibitem{fredman84dict}
Michael~L. Fredman, J\'{a}nos Koml\'{o}s, and Endre Szemer\'{e}di.
\newblock Storing a sparse table with {0(1)} worst case access time.
\newblock {\em Journal of the ACM}, 31(3):538--544, 1984.
\newblock See also FOCS'82.

\bibitem{hagerup01perfhash}
Torben Hagerup and Torsten Tholey.
\newblock Efficient minimal perfect hashing in nearly minimal space.
\newblock In {\em Proc. 18th Symposium on Theoretical Aspects of Computer
  Science (STACS)}, pages 317--326, 2001.

\bibitem{kaplan05permhash}
Eyal Kaplan, Moni Naor, and Omer Reingold.
\newblock Derandomized constructions of $k$-wise (almost) independent
  permutations.
\newblock In {\em Proc. 9th International Workshop on Randomization and
  Approximation Techniques in Computer Science (RANDOM)}, pages 354--365, 2005.

\bibitem{mortensen05dyn1d}
Christian~Worm Mortensen, Rasmus Pagh, and Mihai P\v{a}tra\c{s}cu.
\newblock On dynamic range reporting in one dimension.
\newblock In {\em Proc. 37th ACM Symposium on Theory of Computing (STOC)},
  pages 104--111, 2005.

\bibitem{pagh01redundancy}
Rasmus Pagh.
\newblock Low redundancy in static dictionaries with constant query time.
\newblock {\em SIAM Journal on Computing}, 31(2):353--363, 2001.
\newblock See also ICALP'99.

\bibitem{raman02perfhash}
Rajeev Raman, Venkatesh Raman, and S.~Srinivasa Rao.
\newblock Succinct indexable dictionaries with applications to encoding $k$-ary
  trees and multisets.
\newblock In {\em Proc. 13th ACM/SIAM Symposium on Discrete Algorithms (SODA)},
  pages 233--242, 2002.

\bibitem{raman03succinct}
Rajeev Raman and S.~Srinivasa Rao.
\newblock Succinct dynamic dictionaries and trees.
\newblock In {\em Proc. 30th International Colloquium on Automata, Languages
  and Programming (ICALP)}, pages 357--368, 2003.

\bibitem{schmidt95chernoff}
Jeanette~P. Schmidt, Alan Siegel, and Aravind Srinivasan.
\newblock Chernoff-{H}oeffding bounds for applications with limited
  independence.
\newblock {\em SIAM Journal on Discrete Mathematics}, 8(2):223--250, 1995.
\newblock See also SODA'93.

\bibitem{siegel04hash}
Alan Siegel.
\newblock On universal classes of extremely random constant-time hash
  functions.
\newblock {\em SIAM Journal on Computing}, 33(3):505--543, 2004.

\end{thebibliography}

\end{document}